\documentclass[twocolumn,prl,superscriptaddress,showpacs]{revtex4-1}
\usepackage{graphicx}

\usepackage{color}

\newcommand{\achem}{Anal. Chem.}
\newcommand{\aciee}{Angew. Chem. Int. Ed. Engl.}

\newcommand{\bc}{Biochem.}
\newcommand{\bpj}{Biophys. J.}
\newcommand{\cosb}{Curr. Op. Struct. Biol.}
\newcommand{\cossms}{Curr. Op. Solid State Mat. Sci.}
\newcommand{\cphyc}{Chem. Phys. Chem.}
\newcommand\jsp{J. Stat. Phys.}
\newcommand{\mm}{Macromol.}
\newcommand{\molp}{Mol. Phys.}
\newcommand{\nmat}{Nature Mater.}
\newcommand{\plosone}{PLoS ONE}
\newcommand{\pnasu}{Proc. Natl. Acad. Sci. USA}
\newcommand{\sci}{Science}

\begin{document}

\title{Controlling the folding and substrate-binding of proteins using polymer brushes}

\author{Brenda M.~Rubenstein}
\affiliation{Department of Chemistry, Columbia University, MC 3178, 3000 Broadway, New York NY10027}

\author{Ivan Coluzza*}
\affiliation{Department of Physics, University of Vienna, Boltzmanngasse 5, 1090 Vienna, Austria}

\author{Mark A.~Miller*}
\affiliation{University Chemical Laboratory, Lensfield Road,
Cambridge CB2 1EW, United Kingdom}


\date{\today}

\begin{abstract}
The extent of coupling between the folding of a protein and its binding to a substrate
varies from protein to protein.  Some proteins have highly structured native states in solution,
while others are natively disordered and only fold fully upon binding.  In this Letter,
we use Monte Carlo simulations to investigate how disordered polymer chains grafted around a
binding site affect the folding and binding of
three model proteins.  The protein that approaches the substrate fully folded is
more hindered during the binding process than those whose folding and binding are cooperative.
The polymer chains act as localized crowding agents and can select correctly folded
and bound configurations in favor of non-specifically adsorbed states.  The free energy
change for forming all intra-protein and protein--substrate contacts
can depend non-monotonically on the polymer length.
\end{abstract}

\maketitle

While some globular proteins have a well-defined three-dimensional structure
in solution,
it is now clear that many other proteins possess some degree of intrinsic
structural disorder and that there is a wide variety of reasons why
such disorder may be an advantage for biological function \cite{Dunker02a}.
In many cases, proteins achieve greater order when they bind, thereby
coupling the processes of folding and binding \cite{Wright09a}.
A loss of conformational entropy upon binding
allows the size of the binding free energy to be controlled, and a flexible
unbound structure opens up the possibility of binding to more than one target.
Sometimes, however, the presence of disordered chains is directly linked
to function.
For example, natively unfolded chains in the nuclear pore complex collectively
resemble a polymer brush that provides an entropic barrier to
transport across the nuclear envelope \cite{Lim07a}.
\par
Non-specific interactions with polymer chains have also been employed in the
last fifteen years to create materials whose surfaces resist adhesion of
biological molecules \cite{Kingshott99a,Lai10b}.  Typically, poly(ethylene oxide)
chains are grafted with high density to the surface to provide a steric barrier
to the approach of proteins or larger objects such as cells.
A similar approach can be taken using
oligosaccharide chains, mimicking the non-adhesive role of the glycocalyx in
some cell membranes \cite{Holland98a}.  A small number of theoretical \cite{Halperin11a}
and experimental \cite{Yoshizako02a-short} studies have also considered the effect of
grafted polymers on specific protein--ligand binding.
\par
In this Letter, we investigate how the folding of a protein and its binding
to a substrate can be controlled using a brush of disordered chains surrounding
the substrate.  The chains are grafted to a surface around the
binding site, providing localized steric competition for the protein.  In particular,
we would like to know how the nature of this competition
depends on the extent to which folding and binding are independent or coupled.
\par
Since we are interested in the generic effect of disordered polymer
chains near a binding site, rather than in the details of a specific case,
it is advantageous to employ a coarse-grained model that captures the
essential ingredients of specificity, excluded volume and the entropy of
chains.  The proteins were modeled as strings of amino acid residues that occupy
adjacent sites on a cubic lattice.  A 20-letter alphabet of residues
was used, with interactions between non-bonded residues on adjacent
lattice sites given by the Miyazawa--Jernigan matrix \cite{Miyazawa85a}
and expressed in terms of a reference thermal energy $k_B T_{\rm ref}$,
where $k_B$ is Boltzmann's constant.
Empty lattice sites are taken to represent water molecules, and the interaction
matrix includes an effective treatment of the resulting hydrophobicity of the
relevant residues.
It has recently been shown that principles similar to those of such lattice
models can reproduce accurate protein structures in highly coarse-grained but
off-lattice models \cite{Coluzza11a}.
\par
\begin{figure}
\includegraphics[width=80mm]{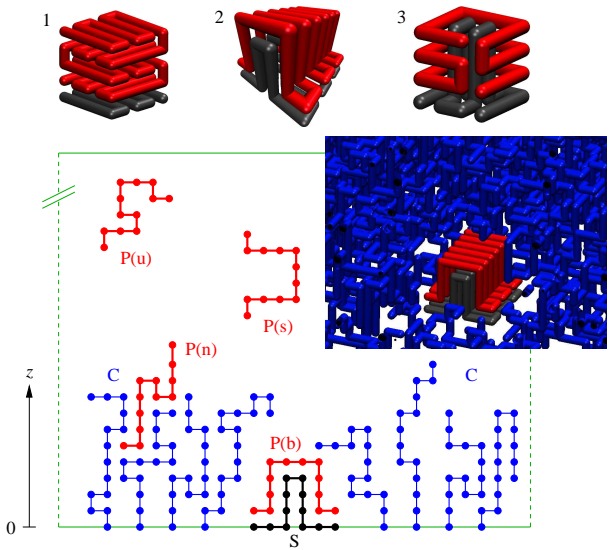}
\caption{
(Color online) Schematic simulation layout.  The protein (P) can be
structured (s) or unstructured (u) in solution and binds
specifically (b) to the substrate (S).  Polymer chains (C) grafted around
the binding site interact non-specifically (n) with the protein.  Periodic
boundary conditions apply in the lateral directions (cell size 30 units).  The
total height of the simulation cell is 150 units.  The three fully bound
protein--substrate geometries are shown above the schematic and the inset shows
a snapshot of protein 2 with polymers.
\label{schematic}
}
\end{figure}

The substrate was constructed from a rigid geometrical arrangement of residues
from the same alphabet as the protein, fixed to the $z=0$ boundary of the simulation
cell.  The grafted polymers were treated as lattice chains with one end fixed
at $z=0$.  These ends were placed on a square grid with a
spacing of three lattice units in each direction and a narrow
ungrafted border around the
binding site.  The polymers interact non-specifically
with each other and with the protein, {\it i.e.}, the interaction is only through
the excluded volume of occupied lattice sites with no energetic contribution.
The simulation box was large enough to allow unfolding of the protein into an
extended coil when unbound and far from the grafted polymers.
The schematic in Fig.~\ref{schematic} illustrates the various broadly defined
states in which the protein may be found.

The protein and substrate sequences were designed by first selecting the folded
and bound conformation of the complex and then applying a Monte Carlo (MC) algorithm
in sequence space \cite{Coluzza04a}.  Trial changes of a residue at a given site were
accepted with the Metropolis probability ${\rm min}[1,\exp(-\Delta E/k_B T_{\rm d})]$,
where $\Delta E$ is the change in energy caused by the residue change and $T_{\rm d}$
is a fixed ``design temperature.''  To prevent the sequence from becoming too
homogeneous (and therefore non-specific), a further acceptance probability of
${\rm min}[1,(N'_{\rm p}/N_{\rm p})^{T_{\rm p}/T_{\rm ref}}]$ was applied, where
$N_{\rm p}=N!/(n_1!n_2!\dots n_{20}!)$ is the number of distinguishable residue
permutations before the trial change and $N'_{\rm p}$ is that afterwards.
$N$ is the number of residues in the complex and $n_i$ the number of residues of
amino acid $i$.
$T_{\rm p}$ is a fictitious ``compositional temperature'' and parallel tempering
was applied in this parameter to enhance exploration of sequence
space.  The replica with the lowest $T_{\rm p}=T_{\rm ref}/14$
equilibrates to a sequence whose
energy is near optimal for the selected conformation \cite{Coluzza04a}.  Folding
can be verified by a conventional MC simulation in configuration space.
\par
The extent of coupling between folding and binding can be influenced before the
sequence design by the choice of the complex's geometry.
For example, a large fraction of intermolecular
contacts tends to make folding conditional on binding, since the
protein--substrate contacts will be required for energetic stability.  During the
design process the coupling can be further controlled by selecting which
interactions to include in the energy change $\Delta E$ of the acceptance
criterion \cite{Coluzza07a}.  Including only interactions across the interface
will again encourage coupled folding and binding, while design of the substrate
after independent design of the protein leads to folding away from the substrate
followed by a lock-and-key binding mechanism \cite{Koshland94a}.
The ability to fold and the coupling
between folding and binding can be checked by ordinary simulation of the designed
sequences on the three-dimensional lattice.  Table~\ref{sequences} gives the
sequences of the three protein--substrate combinations designed for this work.
Protein 1 can attain its folded structure in solution without binding, while protein
2 only gains significant order in the process of binding.  Protein 3 also has
coupled folding and binding but is smaller than protein 2 \cite{supplemental}.
The cooperative folding and binding of proteins 2 and 3 is promoted by large,
non-planar protein--substrate interfaces.
Using $R_g=0.44L^{0.588}$ \cite{Li95a} to estimate the self-avoiding radius of gyration of the 
three fully unfolded proteins, we obtain $R_g\approx6.6$, $5.4$ and $3.6$, respectively.
However, in solution, protein 1 tends to adopt its folded $5\times5\times4$ cuboidal structure
with approximate effective radius $5/\sqrt{2}\approx3.5$.  Hence, proteins 1 and 2 have
comparable unfolded radii, while proteins 1 and 3 have comparable volumes in solution.
The different penetration, folding and binding properties of the three examples are
intrinsically linked to their different geometries and sizes.

\begin{table}
\caption{Designed sequences of the three protein--substrate complexes in
the Miyazawa--Jernigan model \cite{Miyazawa85a}.  A colon separates the
protein sequence from the substrate.
\label{sequences}
}
\begin{tabular}{p{12mm}p{70mm}}
\hline
System & Sequence \\
\hline
1 & {\tt FGCLILWHDGEKDMFPPKEKVRDQAYQMFVCMWRPRERPCFR
         EKDVEKDFTGCCVMWHDREKDMWNPKEKLRDYHYNMWACMWN
         HSEHPCGREKTIEKQG:AYGIAIMWQNSQTYCTSQHTINSSL} \\
2 & {\tt PGTKNKCPCLWTIMICYCENEDGQCFRKNKDHDLWLVMFRYR
         ENEDFCPLRKNKEPDHWIHMNRYRENDDIQ:QGGSSECVMYM
         HLWSWLCKSTITPCFYFQERVMSMFVWAWGDKHFTGHQIAIT
         EKYAGPAYWAVSQKRVALP} \\
3 & {\tt KEHGHGPMDLDEKRIRWYFCTCKERECACMPMQLQE:DNYSN
         AYCCKEKTHRVKDPWMFVGQWSI} \\
\hline
\end{tabular}
\end{table}

Despite the simplicity of the model, it is
computationally demanding to sample thoroughly the folded, unfolded, bound and
unbound conformations of the protein in the presence of the grafted polymers.
In addition to the standard corner-flip, branch rotation, crankshaft and translation
MC moves for the protein and polymer chains \cite{Verdier62a},  configuration bias
MC was used to regrow the grafted polymers \cite{cbmc}.  To improve the ergodicity of
the simulations, a flat-histogram biasing potential \cite{Berg92a} in the energy was built
up using virtual move parallel tempering \cite{Frenkel04a,Coluzza05a}.  The efficiency
of the latter technique comes from its exploitation of information obtained in both
rejected and accepted trial moves.
\par
Using this combination of MC methods we have calculated the free energy profiles
$F(Q)=F_{Q0}-kT\ln P(Q)$ and $F(Z)=F_{Z0}-kT\ln P(Z)$
of the protein--substrate--polymer systems.
Here, $Z$ is the distance of the protein's central monomer from the brush anchoring surface.
$Q$ is the total number of intra-protein and protein--substrate native contacts,
{\it i.e.}, the number of contacts found in the designed structure that
have actually formed, and measures progress from the unbound random coil to the fully
folded and bound state of the protein.
$P(Q)$ and $P(Z)$ are the probability distribution of the order parameters and
$F_{Q0}$ and $F_{Z0}$ are arbitrary free energy offsets that have been chosen
to fix the free energy of the bulk-like protein states for each system to aid comparisons.
Further insight can be obtained by splitting $F(Q)$ into
contributions $F_{\rm ct}(Q)$ and $F_{\rm nc}(Q)$ from states that do and do not
involve contact between
the protein and the substrate.  In the latter case, $Q$ equals the number of intra-protein
contacts.
Although a small number of order parameters can never fully characterize a complex
system, $Q$ and $Z$ are an informative and complementary combination.  We note that the
protein can be unfolded (small $Q$) even when close to the substrate (small $Z$).
For most results, a reduced temperature \cite{Miyazawa85a} of
$T^*=T/T_{\rm ref}=0.3$ was chosen,
where the bound and unbound states of the three proteins
both have significant statistical weights.
\par
\begin{figure}
\includegraphics[width=80mm]{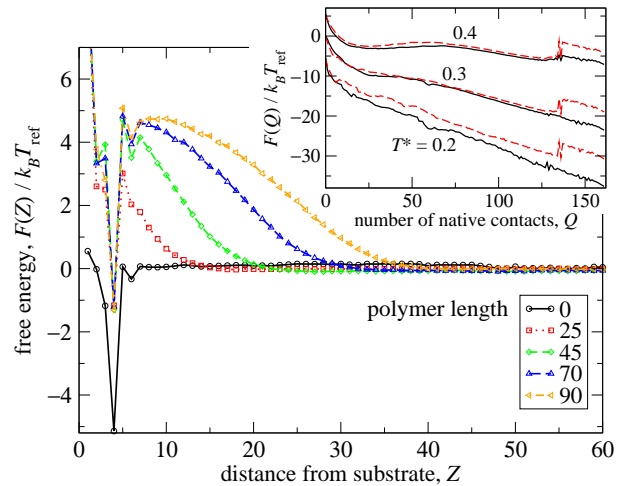}
\caption{
Free energy $F(Z)$ of protein 1
with grafted polymers of length $L$ (see legend), at temperature $T^*=0.3$.  Inset:
free energy $F(Q)$ at three temperatures (marked) without grafted polymers
(solid lines) and with polymers of length $L=40$ (dashed lines).  The data for each
temperature are shifted differently on the vertical axis for clarity.
\label{protein1}
}
\end{figure}

Figure \ref{protein1} shows $F(Z)$ and $F(Q)$ for protein 1.  In the absence
of grafted polymer, $F(Q)$ is steadily downhill at $T^*=0.3$.  The separate contribution
of $F_{\rm nc}(Q)$ (not shown) follows the overall $F(Q)$
closely, but stops abruptly at $Q=136$, which is the
total number of native intra-protein contacts.  Hence, protein 1 readily folds away
from the substrate.  In the presence of polymers of length $L=40$,
$F(Q)$ exhibits a sudden rise at $Q=136$,
which can be attributed to
the loss of polymer entropy due to the fact that the protein must touch the substrate,
disrupting the brush, to gain $Q>136$.
(The small oscillation is due to there being
no unbound states with $Q=135$.)
This lock-and-key protein attempts to
penetrate the brush in its bulky folded state.  At $T^*=0.4$,
the flatter profile of $F_Q$ indicates partial unfolding of the protein in solution
and weaker thermodynamic driving force towards binding.

In the absence of grafted polymers, the $F(Z)$ profile shows a simple,
sharp dip to $F(Z)=-5.1k_B T_{\rm ref}$ on binding.
We can account quite well for this binding free energy simply from
the binding potential energy of the pre-folded protein and the loss of
translational entropy given by the logarithm of the container volume,
yielding $\Delta F\approx-5.5k_B T_{\rm ref}$.
The $F(Z)$ profiles clearly show the barrier introduced by grafted polymers.
The height of the barrier saturates
at around $5k_B T_{\rm ref}$, but its width continues to grow with polymer length.
The bound state of the protein is destabilized by a consistent $4k_B T_{\rm ref}$.
\par
\begin{figure}
\includegraphics[width=80mm]{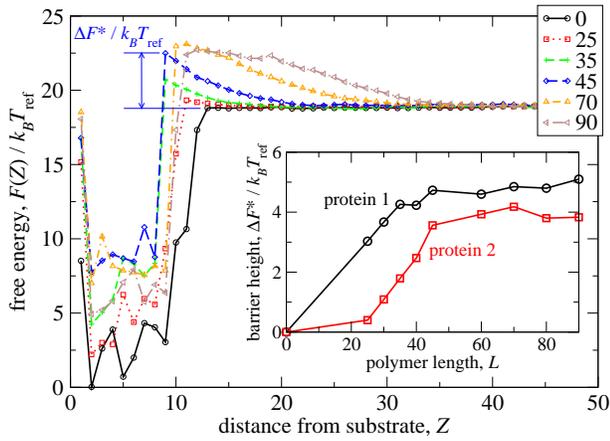}
\caption{
Free energy $F(Z)$ of protein 2
in the presence of grafted polymers of length $L$ (see legend).
Inset: height $\Delta F$ of the barrier experienced by proteins 1 and 2 on approach to the
substrate from large $Z$ due to the polymers.
\label{protein2}
}
\end{figure}

Protein 2 was designed to gain significant native structure only upon binding
with its substrate.
Even without polymers, its $F(Z)$ profile in Fig.~\ref{protein2} shows a
wider well width than the corresponding sharp dip for protein 1 in Fig.~\ref{protein1}.
This is a signature of the protein's relative lack of structure as it approaches the
substrate; in an unfolded state, one part of the chain may interact with the substrate
even while the central monomer is some distance away.
Protein 2's $F(Q)$ profiles with polymers (not shown) all decrease with
$Q$, without sudden jumps, indicating that the onset of its interactions with the
polymer brush is not as abrupt as for protein 1.
As for protein 1, increasing the length of the grafted polymers introduces a barrier
for protein 2 approaching the substrate from large
$Z$.  However, the height $\Delta F^*$ of the barrier grows much more slowly with
$L$ for protein 2 than for protein 1, as shown in the inset of Fig.~\ref{protein2}.
For fairly short polymers, protein 2 experiences a barrier that is only a small
fraction of that for protein 1.  This may be attributed to
protein 2 only gaining its bulky native structure when it is in contact with the substrate.
It therefore causes less disruption while penetrating the brush.  Hence, relatively
short polymers may be able to discriminate between proteins that are structured in
solution and those whose folding and binding are coupled.

\begin{figure}
\includegraphics[width=80mm]{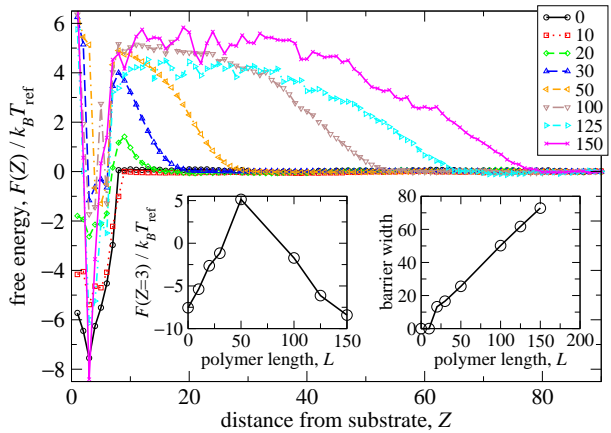}
\caption{Free energy $F(Z)$ of protein 3
with grafted polymers of length $L$ (see legend).
The left- and right-hand insets show how the free energy at binding ($Z=3$) and the
barrier width, respectively, depend on $L$.
\label{protein3Z}
}
\end{figure}

\begin{figure}
\includegraphics[width=80mm]{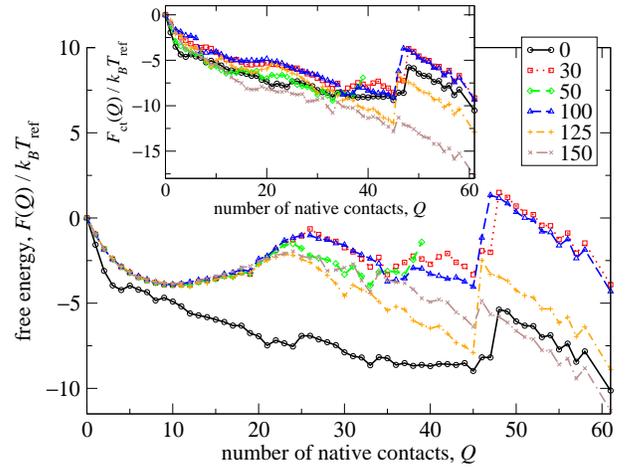}
\caption{
Free energy $F(Q)$ of protein 3
with grafted polymers of length $L$ (see legend).  Inset:
contribution $F_{\rm ct}(Q)$ to the free energy from protein configurations in contact
with the substrate.
\label{protein3Q}
}
\end{figure}

Protein 3 is another sequence that only folds upon binding, wrapping
round a peg-shaped substrate to gain its structure.  Being smaller than protein 2,
it has fewer native energetic contacts to drive its folding and binding.
Its smaller physical extent when unfolded also leads to a slightly narrower
well width in $F(Z)$ (Fig.~\ref{protein3Z}).
As shown by the solid black line in Fig.~\ref{protein3Q}, the free energy $F(Q)$
of protein 3 decreases with increasing contacts $Q$ in the absence of grafted
polymers until $Q=47$.  To reach higher $Q$, a loose end of the protein
must adhere to the substrate, sacrificing entropy.  In the presence of grafted
polymers, states that touch the substrate are entropically disfavored, and $F(Q)$
is dominated by unbound, partially folded states.  Since most such states correspond
to the free protein away from the brush, the $F(Q)$ profiles coincide at low $Q$.  The effect of
removing this contribution is shown in the plot of $F_{\rm ct}(Q)$ in
the inset of Fig.~\ref{protein3Q}.  Note from the behavior of $F_{\rm ct}(Q)$
at high $Q$ that long polymers strongly enhance folding of the protein once it is
in contact with the substrate, relative to the case without polymers, thereby
suppressing partly adsorbed states.

A striking feature of the $F(Q)$ profiles in Fig.~\ref{protein3Q} is the
non-monotonic dependence
of the overall free energy change for folding and binding, $\Delta F=F(60)-F(0)$,
on the polymer length $L$.  As expected, increasing $L$ from 0 initially makes
$|\Delta F|$ smaller because of the loss of brush entropy on protein binding.  This
effect is strong, and we were unable to sample $F(Q)$ beyond $Q=39$ for $L=50$.
However, for larger $L$, the trend is reversed, with $|\Delta F|$ now increasing with
$L$.  Inspection of configurations shows that for intermediate values of $L$,
the free ends of the grafted polymers are able to collapse onto the binding site,
providing an effective steric block.
In contrast, longer polymers force each other to stand more upright by
mutual crowding, thereby releasing space around the binding site itself.  The protein
must still penetrate the brush to reach the site but, once bound and folded,
it does not disturb long polymers as much as short ones, causing the free energy of
binding to become more favorable.  This non-monotonic effect can also be seen in the
$F(Z)$ profile in Fig.~\ref{protein3Z}.  The left-hand inset shows a clear maximum
in the free energy of the bound protein, relative to solution (large $Z$).
The physical width of the barrier (measured at 10\% height) follows the same
monotonic (and rather linear) pattern as the other proteins, as shown in the
right-hand inset of Fig.~\ref{protein3Z}.

We expect two important conclusions of the present work to hold generally.
Firstly, polymer chains around a binding site can assist specific binding by
raising the relative free energy of competing states that involve significant
non-specific adsorption.
Secondly, a polymer brush may be used to introduce a tunable barrier to
binding, thereby allowing control of binding and unbinding rates.  Furthermore,
the barrier can be selective, here acting more strongly on a
protein whose solution structure is well defined than on two examples
that fold and bind cooperatively.  The length of the polymers
can have subtle, non-monotonic results on the effect of the brush.  Altering
the grafting density or polymer topology \cite{Sofia98a} would allow further tuning
of the response.  Finally, we note that the non-specific interactions of tethered
polymers can be useful for controlling non-biological self-assembling
systems \cite{Akcora09a-short}, for example by providing tunable kinetic barriers
to the association of chosen components.
\par
The authors thank the Churchill Foundation of the United States (BMR) and
EPSRC, U.K. (MAM) for financial support.


%

\end{document}